\documentclass[letter,11pt]{article}
\pdfoutput=1 
\usepackage{jheppub} 
\usepackage{caption} 
\usepackage{subcaption}
\usepackage[latin9]{inputenc}
\usepackage{float}
\usepackage{amsmath}
\usepackage{amssymb}
\usepackage{graphicx}
\usepackage{esint}
 \usepackage{hyperref}
\usepackage{comment}
\usepackage{color}
\usepackage{microtype}
\usepackage{cleveref}
\usepackage{breakurl}
\usepackage{bbm}
\newcommand{\be}{\begin{equation}}
\newcommand{\ee}{\end{equation}}
\newcommand{\ben}{\begin{displaymath}}
\newcommand{\een}{\end{displaymath}}
\newcommand{\bea}{\begin{eqnarray}}
\newcommand{\eea}{\end{eqnarray}}
\def\K{K{\"a}hler }
   \newcommand{\rf}[1]{(\ref{#1})}

\def\be{\begin{equation}}
\def\ee{\end{equation}}
\def\bea{\begin{eqnarray}}
\def\eea{\end{eqnarray}}
\def\ba{\begin{array}}
\def\ea{\end{array}}
\def\bit{\begin{itemize}}
\def\eit{\end{itemize}}

\def\rmi{{\rm i}}
\newcommand{\N}{\mathcal{N}}
\newcommand{\cN}{\mathcal{N}}

 \makeatletter

\allowdisplaybreaks

\DeclareRobustCommand{\rcite}[1]{%
  \rcite@aux#1,\@nil{#1}%
}
\def\rcite@aux#1,#2\@nil#3{%
  \if\relax#2\relax
    Ref.~\cite{#3}%
  \else
    Refs.~\cite{#3}%
  \fi
}
\makeatother

\hypersetup{
    colorlinks = true,
    citecolor = {blue},
    linkcolor = {blue},
    urlcolor = {blue},
}

 \title{\rm {\bf \huge  \boldmath Uplifting Anti-D6-brane}}

\author[a]{Niccol\`o Cribiori,}  
\author[b]{Renata Kallosh,}
\author[a]{Christoph Roupec}
\author[a]{and Timm Wrase}
\affiliation[a]{Institute for Theoretical Physics, TU Wien,\\ Wiedner Hauptstrasse 8-10/136, A-1040 Vienna, Austria}
\affiliation[b]{Stanford Institute for Theoretical Physics and Department of Physics,\\ Stanford University, Stanford, CA 94305, USA}
\emailAdd{niccolo.cribiori@tuwien.ac.at}
\emailAdd{kallosh@stanford.edu}
\emailAdd{christoph.roupec@tuwien.ac.at}
\emailAdd{timm.wrase@tuwien.ac.at}

\abstract{A simple, KKLT-like construction of de Sitter vacua in type IIA string theory is presented in an STU model with guidance from string theory U-duality and with an uplifting anti-D6-brane. In four dimensions the model is reduced to $\N=1$ supergravity with three chiral multiplets, namely $S$, $T$ and $U$, as well as one nilpotent multiplet representing the anti-D6-brane. We briefly discuss also a generalization with seven moduli.}

\begin{document}

\maketitle

\section{Introduction}
Cosmological observations during the last two decades suggest that de Sitter and near de Sitter four-dimensional spacetimes are consistent with the data indicating a current and early universe acceleration. It is, however, notoriously difficult to derive de Sitter vacua from string theory compactified to four dimensions, as well as directly in a standard linearly realized four-dimensional supergravity. 

It has been realized in \cite{Kachru:2003aw} that, within type IIB string theory, de Sitter vacua can be obtained in a two-step procedure. First, using stringy perturbative and non-perturbative contributions to an effective superpotential, one can stabilize the volume of the extra dimensions in a supersymmetric anti-de Sitter vacuum. Secondly, with the help of an anti-D3-brane, one has to uplift this anti-de Sitter vacuum with a negative cosmological constant to a de Sitter vacuum with a positive cosmological constant. In four-dimensional $\N=1$ supergravity, this second step of the KKLT construction associated with the uplifting anti-D3-brane can be conveniently described by a non-linearly realized supersymmetry and a nilpotent multiplet \cite{Ferrara:2014kva, Kallosh:2014wsa, Bergshoeff:2015jxa, Kallosh:2015nia, Garcia-Etxebarria:2015lif, Dasgupta:2016prs, Vercnocke:2016fbt, Kallosh:2016aep, Bandos:2016xyu, Aalsma:2017ulu, GarciadelMoral:2017vnz, Cribiori:2017laj, Aalsma:2018pll, Cribiori:2018dlc, Cribiori:2019hod}.\footnote{Actually, in \cite{ Cribiori:2017laj} an anti-D3-brane uplift is obtained by means of a vector multiplet and a new Fayet--Iliopoulos D-term. Its relation to the nilpotent superfield formulation is discussed, too.} A detailed derivation of the KKLT construction of de Sitter vacua from ten dimensions was presented recently in \cite{Hamada:2018qef,Hamada:2019ack, Carta:2019rhx, Gautason:2019jwq, Kachru:2019dvo}.

Moduli stabilization in type IIA string theory was developed in \cite{Grimm:2004ua, Villadoro:2005cu, DeWolfe:2005uu, Camara:2005dc}, but consequently many no-go theorems for de Sitter minima were derived in \cite{Kallosh:2006fm, Hertzberg:2007wc, Haque:2008jz, Flauger:2008ad, Danielsson:2009ff, Wrase:2010ew, Shiu:2011zt, Junghans:2016uvg, Andriot:2016xvq, Junghans:2016abx, Andriot:2017jhf}. A possibility of obtaining metastable de Sitter vacua in type IIA supergravity was proposed in \cite{Kallosh:2018nrk}. It was explained there that, by adding pseudo-calibrated anti-Dp-branes wrapped on supersymmetric cycles, one can generalize the effective four-dimensional supergravity derived from string theory in a way that it includes a nilpotent multiplet. However,  an explicit and simple KKLT-like two-step construction in type IIA string theory has not been presented so far.

In fact, in \cite{Kallosh:2018nrk} an example with an anti-D6-brane is presented which had the following features. The starting point, before the introduction of the anti-D6-brane, is a de Sitter saddle point in linearly realized supergravity. When the action of the anti-D6-brane was added to the system, this saddle point became a de Sitter minimum in which all masses were positive. However, such a de Sitter minimum lied in fact below the de Sitter saddle point, as one can see in Fig.~1 of \cite{Kallosh:2018nrk}. Therefore, this example presenting a stable local de Sitter vacuum in type IIA string theory was of a different nature with respect to the KKLT construction. In particular, the construction in \cite{Kallosh:2018nrk} used only classical ingredients, making an exponentially small cosmological constant unnatural.

Nevertheless, one can take the results of \cite{Kallosh:2018nrk} as a clear indication of the universality of the uplifting nature of pseudo-calibrated anti-Dp-branes wrapped on supersymmetric cycles. The purpose of this paper is therefore to explore constructions involving non-perturbative corrections, that give rise to anti-de Sitter vacua that can in turn be uplifted using anti-D6-branes. Related constructions that reproduce de Sitter vacua in the Large Volume Scenario \cite{Balasubramanian:2005zx} in type IIA have already appeared in \cite{Palti:2008mg}. Here we study constructions similar to the original KKLT model that use only a six-flux and non-perturbative corrections to the superpotential $W$ and an anti-D6-brane uplift.

\section{ The STU model }
\label{sec:STUmodel}

We will employ a simple STU model in order to exemplify how an uplift produced by an anti-D6-brane can be studied in a supergravity setup coming from type IIA string theory. The purpose of this section is therefore to outline such a model and to describe its ingredients.

\subsection{The setup}
In our notation which follows \cite{Dibitetto:2011gm, Danielsson:2013rza}, $S$ is the axio-dilaton, $T$ is a complex structure modulus and $U$ is the volume (K{\"a}hler) modulus.
We propose to use ten-dimensional supergravity compactified on a calibrated manifold, like a Calabi-Yau manifold or a more general SU(3)-structure manifold, such that the standard, linearly realized four-dimensional $\N=1$ supergravity follows. This construction will be supplemented by pseudo-calibrated $\overline {D6}$-branes in order to facilitate a KKLT-like uplift. 

To make our example concrete, we follow \cite{Danielsson:2013rza} and consider a $T^6/(\mathbb{Z}_2\times \mathbb{Z}_2)$ orbifold compactification of type IIA string theory with the ten-dimensional metric 
\be
ds^2_{10} = \tau^{-2}ds^2_4+\rho(\sigma^{-3} G_{ab} dy^a dy^b+\sigma^3 G_{ij}dy^idy^j)\, .
\ee
Here the universal moduli $\rho$, $\tau$ and $\sigma$ are identified as
\be
\begin{aligned}
\rho &= {\rm Im}(U)=(vol_6)^\frac13 ,\\
 \tau &={\rm Im}(S)^\frac14 {\rm Im}(T)^\frac34=e^{-\phi}\sqrt{vol_6},\\
 \sigma  &={\rm Im}(S)^{-\frac16} {\rm Im}(T)^\frac16,
 \end{aligned}
\ee
while $G_{ab}$ and $G_{ij}$ correspond to the two independent three-cycles. To this construction we add $N_{\overline {D6}}^{||}$ and $N_{\overline {D6}}^{\bot}$ anti-D6-branes wrapping three-cycles. The first set of branes extends completely along only one cycle, while the second set corresponds to branes wrapping directions along both cycles, in all the possible combinations.\footnote{See section 2 of \cite{Danielsson:2013rza} for more detailed information on this setup.}

After compactifying to four dimensions, the total scalar potential is given by the sum of two pieces
\be
V_{\rm tot}  = V_{\N=1} + V_{\overline {D6}} \,,
\label{total}\ee
where
\be
\label{eq:VAdS}
V_{\N=1} =e^K\left(g^{i\bar \jmath} D_i W  D_{\bar \jmath} \bar W - 3 W \bar W\right)\,  ,    \qquad  i=\{S,T,U\},
\ee
is the standard $\N=1$ supergravity scalar potential and
\be
\label{eq:vup}
V_{\overline {D6}}= { \mu_1^4  \over {\rm Im} \, (T)^3} + {\mu_2^4   \over  {\rm Im} \, (T)^2  {\rm Im} \, (S)}\,
\ee
is the contribution of the anti-D6-branes. The quantities $\mu_1^4=2 e^{\mathcal{A}_1} N_{\overline {D6}}^{||}$  and $\mu_2^4=2 e^{\mathcal{A}_2}N_{\overline {D6}}^{\bot}$ correspond to anti-D6-branes wrapped on two types of three cycles that are placed in potentially warped regions with warp factors $e^{\mathcal{A}_1}$ and $e^{\mathcal{A}_2}$, respectively.\footnote{For strong warping, the warp factors could in principle depend on the moduli as was the case for the KKLT scenario \cite{Kachru:2003sx}. This could modify our models quantitatively but not qualitatively. To investigate this, it would be interesting to extend the analysis of the anti-D3-brane in the KKLT scenario \cite{Cribiori:2019hod} to our type IIA setup with anti-D6-branes.}

The \K and superpotential for our model are 
\be
\begin{aligned}
K &= - \log \left(-\rmi (S-\bar S)\right) - 3 \log \left(-\rmi (T-\bar T)\right)-3 \log \left(-\rmi (U-\bar U)\right)\,,\\
W &= f_6 + W_{np}\,,
\label{our}\end{aligned}
\ee
where $f_6 $  is the flux parameter for a six-flux and the non-perturbative part $W_{np}$ of the superpotential is
\be
W_{np}  = \sum_i A_i e^{\rmi a_i \Phi_i}, \qquad \qquad \Phi_i=\{S,T,U\}.
\label{np}\ee
We furthermore assume that all of the parameters $f_6$, $A_S$, $A_T$, $A_U$, $a_S$, $a_T$ and $a_U$  are real and constant. Indeed, the parameters $A_i$ can, in principle, depend on the corresponding moduli $\Phi_i$ like $A_i (e^{\Phi_i})  \approx A_i(0) + A^\prime _i (0) e^{-{\rm Im}\Phi_i} + \cdots $. However, for $e^{-{\rm Im}\Phi_i} \ll 1$ only the constant contribution will be significant. We will thus restrict our analysis to this case.

The non-perturbative part of the superpotential \eqref{np}, involving $S$ and $T$, may arise from gaugino-condensation. Indeed, this effect can be parameterized by introducing terms of the form $e^{- a / g_{YM}^2}$ into the superpotential, where $a = \frac{2 \pi}{N}$ and $g_{YM}$ is the coupling constant of the Yang-Mills theory living on $D6^\parallel$ or $D6^\perp$. It is possible to identify the coupling constants with the moduli $S$ and $T$ in the following way \cite{Danielsson:2013rza}:
\be
\frac{1}{\left(g_{YM}^\parallel \right)^2 } \sim {\rm Im} (S),\qquad  \qquad
\frac{1}{\left(g_{YM}^\perp \right)^2 } \sim {\rm Im} (T) \,.
\label{ST}\ee
Alternatively, Euclidean D2-branes wrapping internal 3-cycles will likewise give rise to such terms with $a_S = a_T = 2\pi$. The origin of the non-perturbative term depending on the modulus $U$ will be motivated in the following subsection.

Note that, in principle, the superpotential can have the more generic form
\be\label{eq:Wpert}
W= f_6 + f_4 U + f_2 U^2 + f_0 U^3 + \left( h_T + r_T U\right) T + \left( h_S + r_S U\right) S + W_{np} \,,
\ee
with $f_p$ ($p = 0, 2, 4,6$) arising from RR-fluxes, $h_{S/T}$ from integrating the NSNS-flux over the corresponding 3-cycles and $r_{S/T}$ from the curvature of the internal manifold. However, in complete analogy with the KKLT setup, which has the superpotential $W_{KKLT}=W_0 + A_\rho e^{\rmi a_\rho \rho}$, we keep only $f_6$ and the non-perturbative exponents. Indeed, as we will discuss in appendix \ref{appA} for some examples, the inclusion of flux contributions, with the exception of $f_6$, seems to prohibit our uplift procedure from working. 

There are many no-go theorems forbidding de Sitter vacua when only certain sets of classical ingredients are employed, see for example \cite{Kallosh:2006fm, Hertzberg:2007wc, Haque:2008jz, Flauger:2008ad, Danielsson:2009ff, Wrase:2010ew, Shiu:2011zt, Junghans:2016uvg, Andriot:2016xvq, Junghans:2016abx, Andriot:2017jhf}. Recently it has been proposed that the inclusion of anti-D6-branes \cite{Kallosh:2018nrk, Banlaki:2018ayh} or KK monopoles \cite{Blaback:2018hdo} can evade all no-go theorems and lead to classical, metastable dS vacua. In this work we evade the no-go theorems by including non-perturbative corrections that indeed turn out to be essential. Interestingly, our approach in this paper does not require a non-vanishing Romans mass parameter $f_0$ \footnote{In fact, one of the no-go theorems \cite{Haque:2008jz,Flauger:2008ad} explicitly requires $f_0 \neq 0$ for the possibility of de Sitter vacua at tree level.} and therefore, contrary to the other constructions, it does allow for a direct lift to M-theory.

It is known that the KKLT construction provides a well working mechanism of stabilizing one complex modulus in an anti-de Sitter vacuum, assuming that the other moduli were already stabilized by perturbative terms in the superpotential. We will show that the same mechanism is also working well in each of the three complex moduli directions in our type IIA model. Indeed, we will follow a two-step procedure:
\begin{enumerate}
\item We stabilize all moduli using the six-flux and non-perturbative corrections. In this way we obtain a stable and supersymmetric anti-de Sitter vacuum, in which all of the fields have positive masses.
\item We then uplift the vacuum to de Sitter by adding anti-D6-branes, namely we introduce the corresponding terms with coefficients $\mu_1$ and $\mu_2$ in the scalar potential, as seen in equation \eqref{eq:vup}. 
\end{enumerate}
We notice that, even if the first point is met, in general it is not guaranteed that the anti-de Sitter vacuum can consistently be uplifted to de Sitter. Indeed, examples in which the uplift fails are presented in the appendix \ref{sec:failedmodels}.

\subsection{Satisfying stringy requirements}\label{sec:stringy}
For a consistent embedding of our setup in type IIA string theory, we have to satisfy Gauss' law in the compact space. This amounts to satisfying the Bianchi identities for the RR fields $F_p$. In our case for a compactification without 1- and 5-forms, the only non-trivial Bianchi identity is the tadpole condition for the D6-brane charges. It takes the form
\be\label{eqn:tadpole}
\int dF_2 -F_0 H = -2 N_{O6} + N_{D6} - N_{\overline {D6}}\,
\ee
and needs to be satisfied for each three-cycle independently. Since we are interested in very simple models with only non-vanishing $F_6$ flux, we have to satisfy the tadpole condition by adding D6-branes. These will cancel the negative contributions from the O6-planes and potential anti-D6-branes. In order to avoid instabilities due to the presence of D6-branes and anti-D6-branes, one has to find an appropriate geometry, which might be non-trivial (see for example \cite{Retolaza:2015sta}). Alternatively, together with non-perturbative effects from Euclidean D2-branes, one could try to also include more exotic O6-planes, like anti-O6-planes \cite{Kachru:1999ed} or O$6^+$-planes. In particular, anti-O6-planes  have negative tension but opposite RR-charge and this means that they would contribute to the tadpole condition in equation \eqref{eqn:tadpole} with the same sign as D6-branes. Since they are non-dynamical, one would not have to worry about related instabilities. We leave a detailed study of this aspect to the future.

The RR- and NSNS-fluxes also have to be appropriately quantized. For us this means in particular that the parameters $f_p$ in the superpotential in equation \eqref{eq:Wpert} can assume only discrete values. We can easily set the $f_6$ parameter in equation \eqref{our} to any particular value by rescaling the superpotential. Indeed, this will change $f_6$ and the $A_i$ in \eqref{np} but neither the location nor the existence of the vacua.

Lastly, we would also like to ensure that higher order non-perturbative corrections, as well as $\alpha'$ and string loop corrections, are suppressed. The general form of the non-perturbative corrections of the kind we are considering is a sum over all the instanton contributions
\be
\label{eq:fullnp}
\sum_{n=1}^\infty A_n e^{\rmi n a_i \Phi_i}
\ee
for all the fields $i=\{S,T,U\}$. However, in the non-perturbative superpotential in equation \eqref{np}, for each of the moduli we are keeping only the very first term in the sum \eqref{eq:fullnp}. To consistently neglect all the $n>1$ terms it is necessary to require that  $a_i \text{Im}(\Phi_i) > 1$, $\forall i$. Additionally, in order to suppress $\alpha'$ corrections and trust the supergravity approximation we need to require the volume of the internal manifold to be large, i.e. $vol(6)\gg 1$. Since in our setup $vol(6)=({\rm Im}(U))^3$, we will demand Im$(U) \gg 1$. Finally, string loop corrections are expected to be suppressed if Im$(S)\gg \tfrac{1}{16\pi^2}$. As  we discuss in the next subsection, all of these requirements can be satisfied by using scaling symmetries of the STU model.

\subsection{Scaling properties of the STU Model} 

From the previous discussion it seems that stringy requirements restrict the allowed positions of the critical points of the scalar potential. However, it is known since the racetrack inflation model \cite{BlancoPillado:2004ns} that one can obtain models with rescaled values of the critical points for different choices of parameters, with respect to the original ones.

 \begin{itemize}
 \item
 In our STU model we notice that the kinetic terms are invariant under the rescaling
\be
S \rightarrow \lambda_S S\, , \qquad T \rightarrow \lambda_T T\, , \qquad U \rightarrow \lambda_U U.
\ee
Indeed the \K potential changes only by an additive constant, namely $-\log(\lambda_S\lambda_T^3\lambda_U^3)$, which can be compensated with a \K transformation. 
The resulting modification of the superpotential, then, can then be taken into account by rescaling of the parameters $a_i$
\be
a_S \rightarrow a_S/\lambda_S\, , \qquad
a_T \rightarrow a_T/\lambda_T\, , \qquad  a_U \rightarrow a_U/\lambda_U,
\ee
and by sending also $\{f_6,A_i\}\to (\lambda_S\lambda_T^3\lambda_U^3)^\frac12\{f_6,A_i\}$. Altogether, this implies that we can in fact rescale the positions of the anti-de Sitter minima from one set of fields $S$, $T$ and $U$, associated to a given choice of $a_i$, to another set with rescaled $a_i$ parameters. In the case in which the uplifting term is present, we also need to rescale 
\be
\mu_1^4 \rightarrow \mu_1^4 \lambda_T^3\, , \qquad \mu_2^4 \rightarrow \mu_2^4 \lambda_T^2 \lambda_S\,
\ee
and send the nilpotent chiral multiplet $X\to(\lambda_S\lambda_T^3\lambda_U^3)^\frac12 X$. This last ingredient will be introduced in section \ref{sec:nil}, in order to describe the anti-D6-brane contribution within an effective supergravity theory.

\item We also notice that an overall rescaling of the scalar potential, namely $V \rightarrow c^2 V $, can be achieved via
\be
f_6 \rightarrow c f_6 \, , \qquad A_i  \rightarrow c A_i  \, , \qquad  \mu_1^4 \rightarrow c^2 \mu_1^4  \, , \qquad  \mu_2^4 \rightarrow c^2 \mu_2^4\,.
\ee
\end{itemize}
As a consequence of these transformations, we can change the parameters in the scalar potential as well as the critical points at which the moduli are stabilized. The presence of the (anti-)de Sitter minimum for the new parameters and critical points of the moduli is guaranteed by the rescaling properties of the theory.

\section{The $U$-exponent issue}

Two possible explanations of the non-perturbative terms in the $S$ and $T$ directions were given in \cite{Palti:2008mg}, where it is argued that they can arise either from gaugino condensation on stacks of D6-branes or from Euclidean D2-branes. We also commented on this fact previously, near equation \eqref{ST}. On the other hand, the origin of the non-perturbative term in the volume modulus $U$ is less clear. We will  discuss this issue in the present section. 

We propose two arguments to justify the presence of exponential terms in $U$ in the non-perturbative superpotential. 
\begin{enumerate}
\item The first reason is the concept/conjecture about string theory U-duality, which follows from M-theory. String theory tends to have S-duality and T-duality, which at the level of M-theory are expected to be combined into U-duality \cite{Hull:1994ys,Schwarz:1996bh}. This is known as a discrete U-duality symmetry, $E_7(\mathbb{Z})$, which contains the S-duality and T-duality groups as subgroup: 
\be
E_7(\mathbb{Z}) \supset SL(2, \mathbb{Z})\times O(6,6; \mathbb{Z})
\ee

To this observation, one can add that the supersymmetric STU black holes, which have a symmetry known as string triality, have played a significant role in studies of the non-perturbative states of string theory, see for example \cite{Behrndt:1996hu}. From this perspective it appears natural to expect that the non-perturbative exponential terms in $W$ are possible not only in the $S$ and $T$ directions, but also in the $U$ directions, since at the level of M-theory these moduli appear on equal footing, see for example \cite{Acharya:2007rc}.

The problem, however, is to identify a specific mechanism in type IIA string theory which is capable of producing the terms $e^{\rmi a_U U}$ in $W$, in addition to $e^{\rmi a_S S}$ and $e^{\rmi a_T T}$ as we define in equation \eqref{np} above. We may suggest that such term may originate from instantons, analogous to an Euclidean D3-brane wrapping a four-cycle in type IIB string theory, \cite{Witten:1996bn,Kachru:2003aw}. The problem here is that in the early studies of string theory instantons very often string theory moduli were viewed as constants, without the need to stabilize them as a function of the four-dimensional spacetime. 

In type IIA string theory  we may think about  instantons from an Euclidean NS5-brane wrapping a six-cycle. However, such instantons in the context of volume stabilization were already studied in \cite{Looyestijn:2008pg}. It was conjectured there that these can only lead to corrections to the \K potential, since the volume itself cannot be expressed as a holomorphic function of the $\cN=1$ chiral superfields.  It would mean that NS5-brane instantons are not useful for our purpose. However, the argument/conjecture in \cite{Looyestijn:2008pg}  is not fully clear and might need to be revisited. 

If we look at  earlier treatment of Fivebrane Instantons in Sec. 4 of \cite{Becker:1995kb}, we may notice that the volume of the six-dimensional manifold is not present in their equations (4.1) and (4.2). The instanton action of the Fivebrane is given by 
\be e^{-S_6} = e^{- {1\over  g_s^2} - {ia} },
\ee 
where $a$ is the axion field. There is no six-dimensional volume in this expression and it is not clear how to apply this to our situation.  However, this form of the instanton action shows clearly that one of the U-duality symmetries, namely the axion shift symmetry of the theory which is part of the $SL(2, \mathbb{R})$ symmetry,
\be
a\rightarrow a+c,
\ee
is broken unless  $c= 2\pi n$, where $n$ is an integer. This identification breaks the continuous $SL(2, \mathbb{R})$ symmetry down to its discrete $SL(2, \mathbb{Z})$  subgroup.

Notice that the other parts of the $E_7(\mathbb{Z}) $ U-duality symmetry mix our three moduli, $S$, $T$ and $U$, with some discrete parameters. To start with supergravity where we have a continuous U-duality and to end up with a discrete one, we need exponential terms involving axions, as we explained above, but we need them in all directions, including $U$.

In the past, it was not always easy to find evidences of stringy U-duality, but many useful examples were found.  Here, we will suggest to consider string wormholes \cite{Giddings:1989bq} as a possible evidence towards U-duality and a possible source of the non-perturbative exponential for the modulus $U$. Indeed, consider the case in which the ten-dimensional theory is reduced to four dimensions by compactification on a six-dimensional Calabi-Yau (or other) manifold and assume that the ten-dimensional metric  has a breathing mode, related to the volume modulus of the six-dimensional space. In this setup, the wormhole solution of \cite{Giddings:1989bq} involves the radial axion and the breathing mode, which form a natural complex variable
$
Z= i U
$.
In particular, the real part of $Z(x) $ is related to a breathing mode field and the imaginary part is an axion field. The kinetic term for the complex field $Z$ was identified in \cite{Giddings:1989bq}  in equation (15). In this equation, we take a more recent standard normalization of the scalar curvature term in supergravity into account, with $1/2 R$ as the Einstein term rather than $ R$. Then the kinetic term of a modulus associated with the stringy wormhole becomes
\be
-3 {\partial Z \partial \bar Z\over (Z+\bar Z)^2}\,.
\ee
This supports the identification of $Z$ with our volume modulus $U$. Thus, we find that the stringy wormholes presented in \cite{Giddings:1989bq} suggest evidence for the existence of a non-perturbative exponential term in string theory associated with the modulus $U$, which in type IIA  string theory describes the volume modulus or, using the earlier name, the breathing mode field of the ten-dimensional metric defined in equation (3) of \cite{Giddings:1989bq}. 

This fact by itself does not tell us that such non-perturbative exponential terms in string theory, associated with the modulus $U$, have to appear in the superpotential $W$. The reasoning here which we suggest is the fact that in type IIA such exponential terms in the $S$ and in $T$ directions do appear. Thus, if U-duality is indeed a property of non-perturbative string theory, we can conclude that we do have a reasonable expectation that non-perturbative exponential terms for the modulus $U$ are possible in $W$. It would be very nice to understand these issues much better.

In addition to this, open string worldsheet instantons in $\N=1$ orientifold compactifications of type IIA string theory are generically  giving rise to the required exponential terms in $W$ that depend on the \K moduli \cite{Kachru:2000ih,Blumenhagen:2009qh}.\footnote{We are grateful to Ralph Blumenhagen for pointing this out to us.}

\item Our second reason to add an exponential in $U$ to the superpotential is the following. In type IIA string theory it was always difficult to find de Sitter vacua and even more difficult to find de Sitter minima. We will show below that with the new ingredients: non-perturbative exponents in $W$ in all directions and uplifting anti-D6-branes, building de Sitter vacua is not complicated anymore. In fact, our new cosmological model with all three exponents, respecting the U-duality symmetry of string theory, works extremely well for the purpose of getting anti-de Sitter minima, which in turn admit an uplifting via anti-D6-branes, producing de Sitter minima.

\end{enumerate}

We can start now with the analysis of the model presented in section \ref{sec:STUmodel}. Indeed, in the following two sections we will show how to construct de Sitter vacua in type IIA string theory.

\section{Supersymmetric anti-de Sitter minimum}
The first step in constructing a KKLT-like de Sitter vacuum is finding a stable anti-de Sitter vacuum of the $\N=1$ supergravity scalar potential \eqref{eq:VAdS}. For stability in an anti-de Sitter spacetime it is sufficient to satisfy the well-known Breitenlohner-Freedman bound, which allows for negative mass values. However, in the present work we prefer to make a stronger request and ask that all the masses are positive. Indeed, we believe that such a situation is preferable in order to avoid instabilities at the step in which the anti-de Sitter vacuum will be uplifted to de Sitter.

To find a supersymmetric anti-de Sitter vacuum it is sufficient to solve the F-term equations
\be
D_i W=0, \qquad i=\{S,T,U\},
\label{susy}\ee 
using the \K potential and the superpotential given in (\ref{our}) and (\ref{np}). It is a known fact that these equations imply $\partial_i V=0$ but, importantly, not the other way around. For simplicity, at the supersymmetric anti-de Sitter minimum we set the axions to zero, namely those fields which receive a mass only through $W_{np}$:
\be
{\rm Re} \, (S)= {\rm Re} \, (T)= {\rm Re} \, (U)=0\,.
\ee
Such an assumption can be safely made as long as the masses in the vacuum are positive. 
Denoting the positions of the remaining fields at the minimum with 
\be
{\rm Im} \, (S)=  S_0 \, ,   \qquad {\rm Im } \, (T)= T_0  \, ,   \qquad {\rm Im } \, (U)= U_0 \,,
\ee
we solve the equations \eqref{susy} for the pre-exponential factors $A_i$ and find an expression for them in terms of the seven parameters of our choice:
\be
\label{eq:AdSsol}
A_i = A_i (f_6, a_S, a_T, a_U, S_0, T_0, U_0)\,, \qquad i=\{S,T,U\}.
\ee
Notice that we are keeping $f_6$ as a free parameter. This choice will make the flux quantization straightforward to implement. Once the solution \eqref{eq:AdSsol} is known, it is possible to check its stability by calculating the canonically normalized mass matrix
\be
{m_i}^j = \frac12 g^{jk}\nabla_k\partial_i V
\ee
at the minimum.\footnote{Here $g_{ij}$ is obtained by rewriting the \K metric in real coordinates.} Notice that, upon substituting \eqref{eq:AdSsol}, the mass matrix becomes a function of the aforementioned seven parameters: $m_{ij} = m_{ij}(f_6, a_S, a_T, a_U, S_0, T_0, U_0)$.

As discussed in subsection \ref{sec:stringy}, there are some restrictions that we need to impose on the general solution \eqref{eq:AdSsol}. A first requirement is related to the non-perturbative corrections. In order to consistently neglect higher order instanton contributions in the superpotential, we will choose the parameters $a_i$ such that $e^{-a_i \text{Im}(\Phi_i)}$ is smaller than $\mathcal{O}(10^{-1})$ for each $i$ individually, i.e. along each field direction. Second, in order to trust the supergravity approximation of string theory, we have to require the volume of the internal manifold to be large. This will be implemented by choosing a sufficiently large value for the parameter $U_0$, $e.g.:$ $U_0\simeq\mathcal{O}(10)$. Finally, we will set the parameter $S_0$ to be $\mathcal{O}(1)$, such that string loop corrections are suppressed.

In the following we show that the class of stable and supersymmetric anti-de Sitter solutions \eqref{eq:AdSsol} with the aforementioned restrictions is not empty. In particular, we will give two concrete examples of such solutions, corresponding to two different choices of the free parameters, and we will check that the masses in the vacuum are all positive. Since we find that the required properties hold, even for small variations of the parameters, we believe that solutions of the type \eqref{eq:AdSsol} have a sufficiently large parameter space and are not isolated points.

\subsection{Two specific anti-de Sitter solutions}

As we will show in this subsection, by giving two explicit examples, it is not hard to find some set of parameters giving a solution of the type \eqref{eq:AdSsol}, with the desired properties. In both cases we set $S_0 = T_0 =1$ for simplicity and we choose $U_0=10$ in order to implement the large volume approximation. We then are left with four parameters, namely $f_6$ and the triplet $a_i$. Two possible choices for them are given in the following table.
\begin{table}[H]
\center
\begin{tabular}{|c|c|c|c|c|}\hline
 & $f_6$ & $a_S$ & $a_T$ & $a_U$ \\\hline
Set 1 & $1$ & $3$ & $3$ & $0.5$ \\\hline
Set 2 &$2$ & $3.1$ & $3.3$ & $0.32$ \\\hline
\end{tabular}
\caption{The two sets of parameters that we investigated in detail. For both of them we find a stable anti-de Sitter vacuum with positive masses.}
\label{tab:para}
\end{table}
Notice that the parameter $a_U$ is roughly one order of magnitude smaller than $a_S$ and $a_T$, but the restriction $e^{-a_U U_0}< 10^{-1}$ is still satisfied since $U_0$ is one order of magnitude bigger than $S_0$ and $T_0$.
We can now check the stability of these two anti-de Sitter vacua by calculating the masses for the various fields. The results are reported in table \ref{tab:massesADS}.
\begin{table}[H]
\center
\begin{tabular}{|c|c|c|c|c|c|c|}\hline
& $m_{1}^{\,2}$ & $m_{2}^{\,2}$  & $m_{3}^{\,2}$ & $m_{4}^{\,2}$ & $m_{5}^{\,2}$ & $m_{6}^{\,2}$ \\\hline
Set 1 & $4.36\cdot10^{-4}$& $3.79\cdot 10^{-4}$& $1.01\cdot 10^{-4}$ & $7.37\cdot 10^{-5}$& $5.66 \cdot 10^{-5}$ & $3.64 \cdot 10^{-5}$ \\\hline
Set 2 & $1.19 \cdot 10^{-3}$ & $1.01\cdot 10^{-3}$& $2.43 \cdot 10^{-4}$ & $2.20 \cdot 10^{-4}$& $1.64 \cdot 10^{-4}$ & $1.45\cdot 10^{-4}$  \\\hline
\end{tabular}
\caption{The canonically normalized masses squared for both sets of parameters are all positive.} 
\label{tab:massesADS}
\end{table}
With the parameters given in table \ref{tab:para}, we get the values for the $A_i$ and $e^{-a_i {\rm Im}\Phi_i}$, listed below in table \ref{tab:As}. To check whether or not these make sense, we need to evaluate $A_i e^{- a_i {\rm Im} \Phi_i}$, which should be smaller than the tree level contribution of $f_6$, as discussed in section \ref{sec:stringy}. This is the case for our values, which were indeed chosen to comply with these requirements.
\begin{table}[H]
\center
\begin{tabular}{|c|c|c|c||c|c|c|}\hline
 & $A_S$ & $A_T$ & $A_U$  & $e^{-a_S {\rm Im} S}$ & $e^{-a_T {\rm Im} T}$ & $e^{-a_U {\rm Im} U}$  \\\hline
Set 1 & $-1.70$ & $-5.11$ & $-22.6$ & $0.0498$ & $0.0498$ & $0.00674$  \\\hline
Set 2 & $-3.43$ & $-11.8$ & $-11.0$ & $0.0450$ & $0.0369$ & $0.0408$  \\\hline
\end{tabular}
\caption{The resulting values for the parameters in front of the exponentials.}
\label{tab:As}
\end{table}

We did not encounter any particular difficulties in finding an appropriate choice for the parameters that yields an anti-de Sitter vacuum with the desired properties. For this reason, it seems plausible that a considerably large parameter space of working models exists. We notice that, when increasing the values of the parameters $a_i$, the masses approach zero. This is expected, since in the regime of large $a_i$ the non-perturbative corrections become small.

\section{ Uplifting anti-D6-branes and de Sitter minimum}
In this section we show how the previously found anti-de Sitter vacua can be uplifted to de Sitter. For this purpose, we add a new ingredient, namely a certain number of anti-D6-branes, to the setup. We stress that, even if an anti-de Sitter vacuum with the desired properties is found, it is not guaranteed that it can be always uplifted to de Sitter. On the contrary, in many situations we looked at the uplift did not work, eventually giving one or more runaway directions. The very fact that we can uplift the two anti-de Sitter vacua presented in the previous section is a non-trivial result.
\begin{figure}[h!]
\includegraphics[scale=0.5]{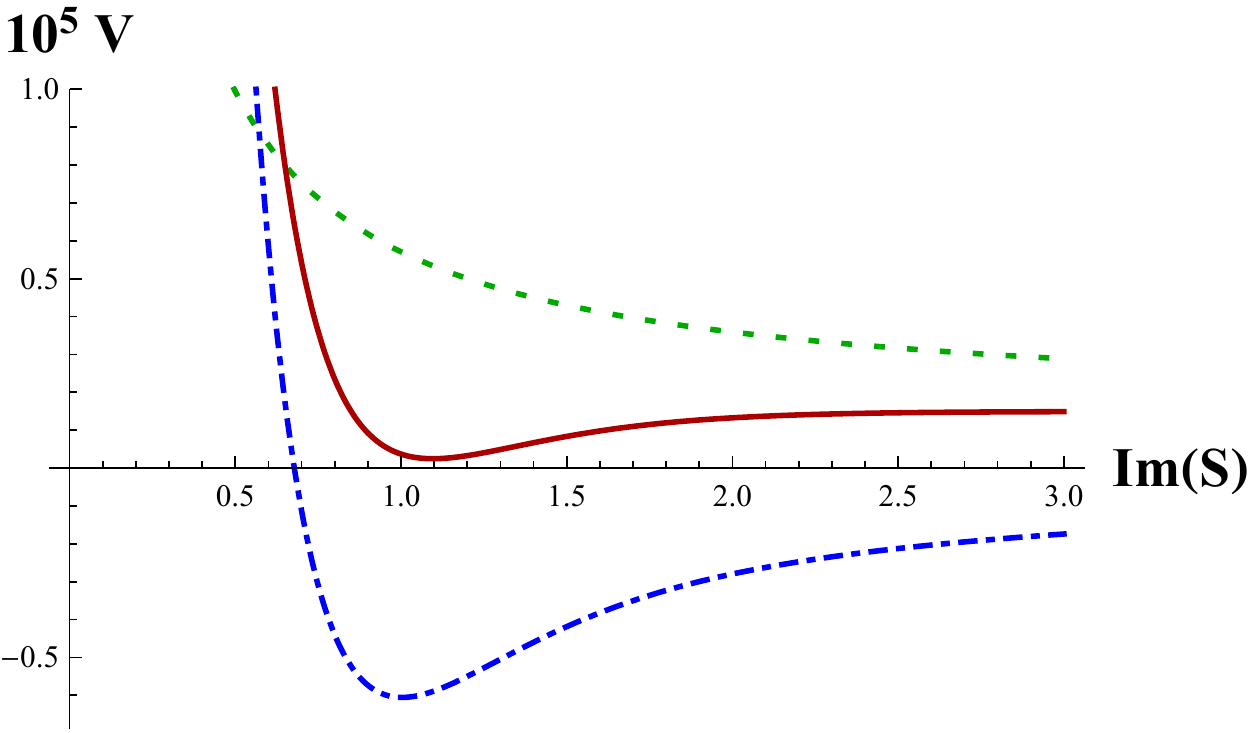}\hspace{10pt} \includegraphics[scale=0.5]{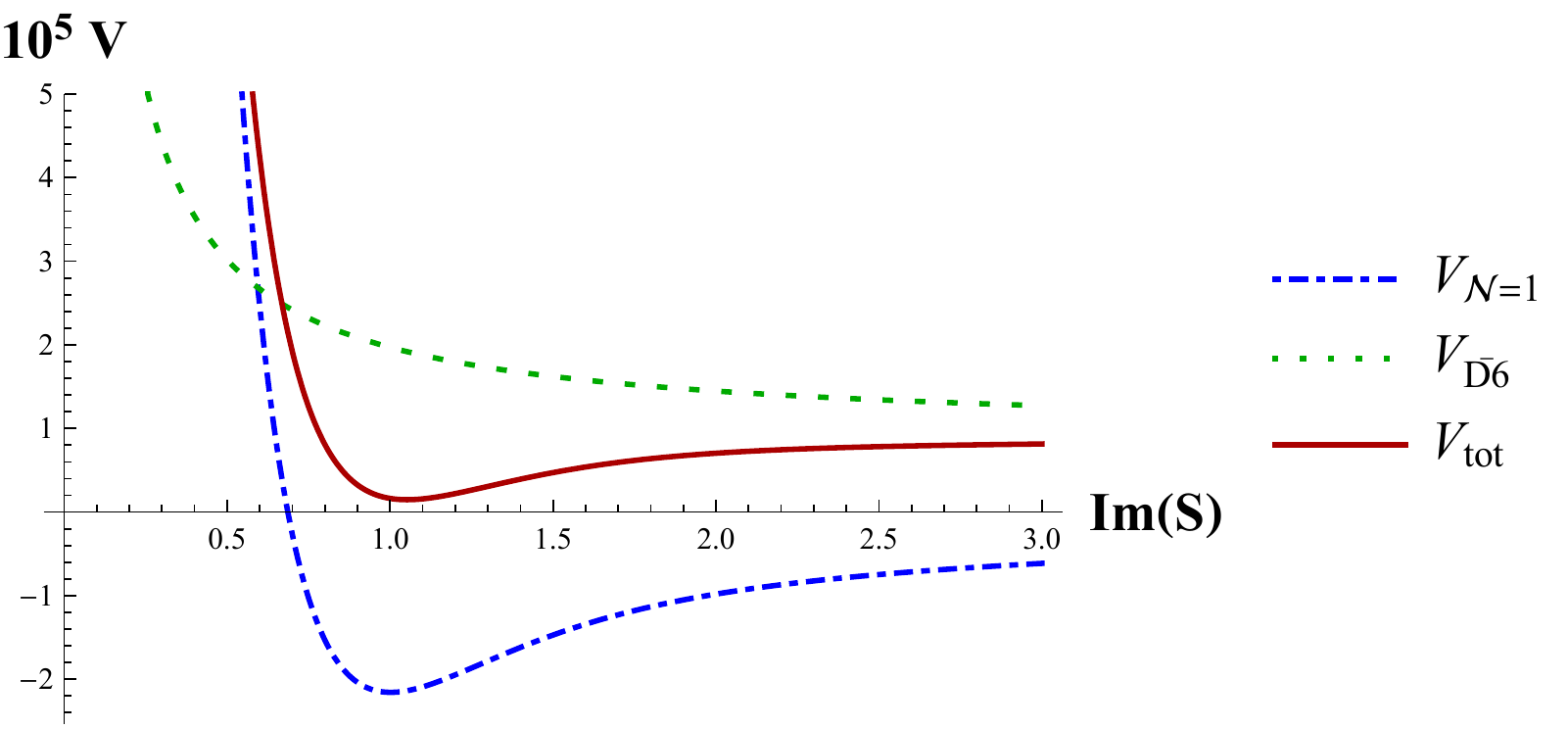}\vspace{15pt}\\
\includegraphics[scale=0.5]{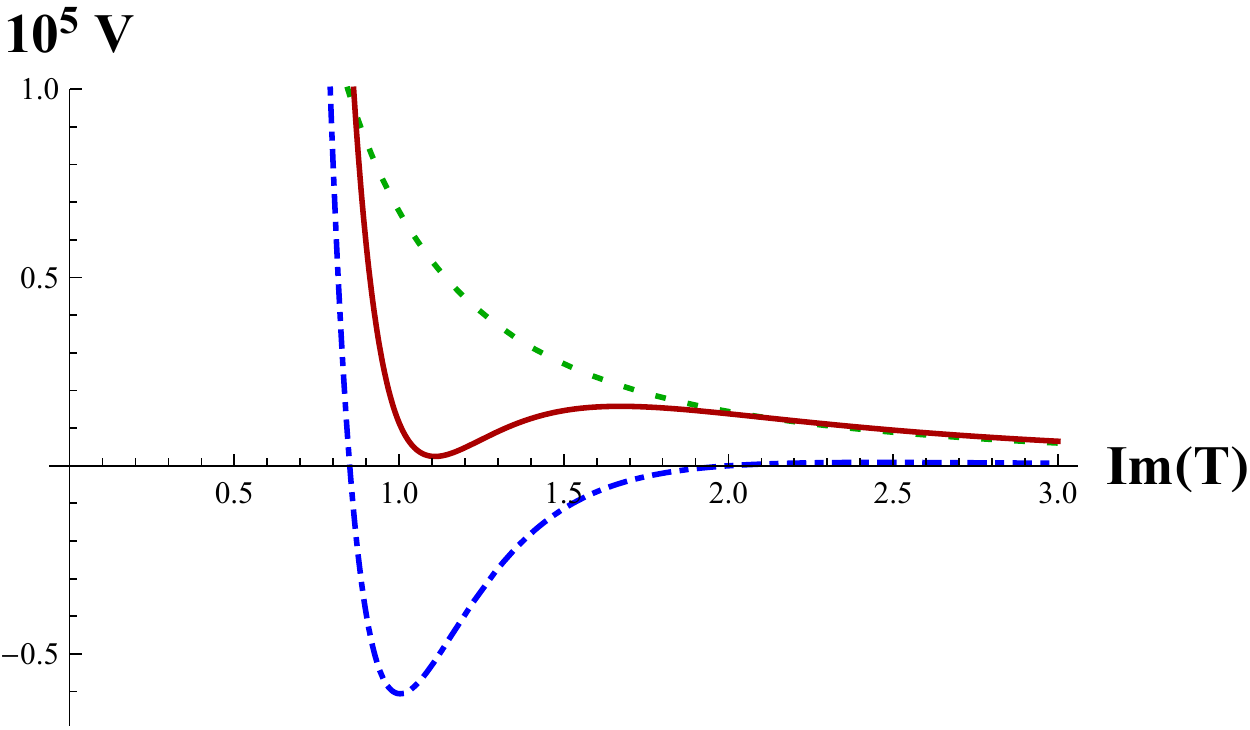}\hspace{10pt} \includegraphics[scale=0.5]{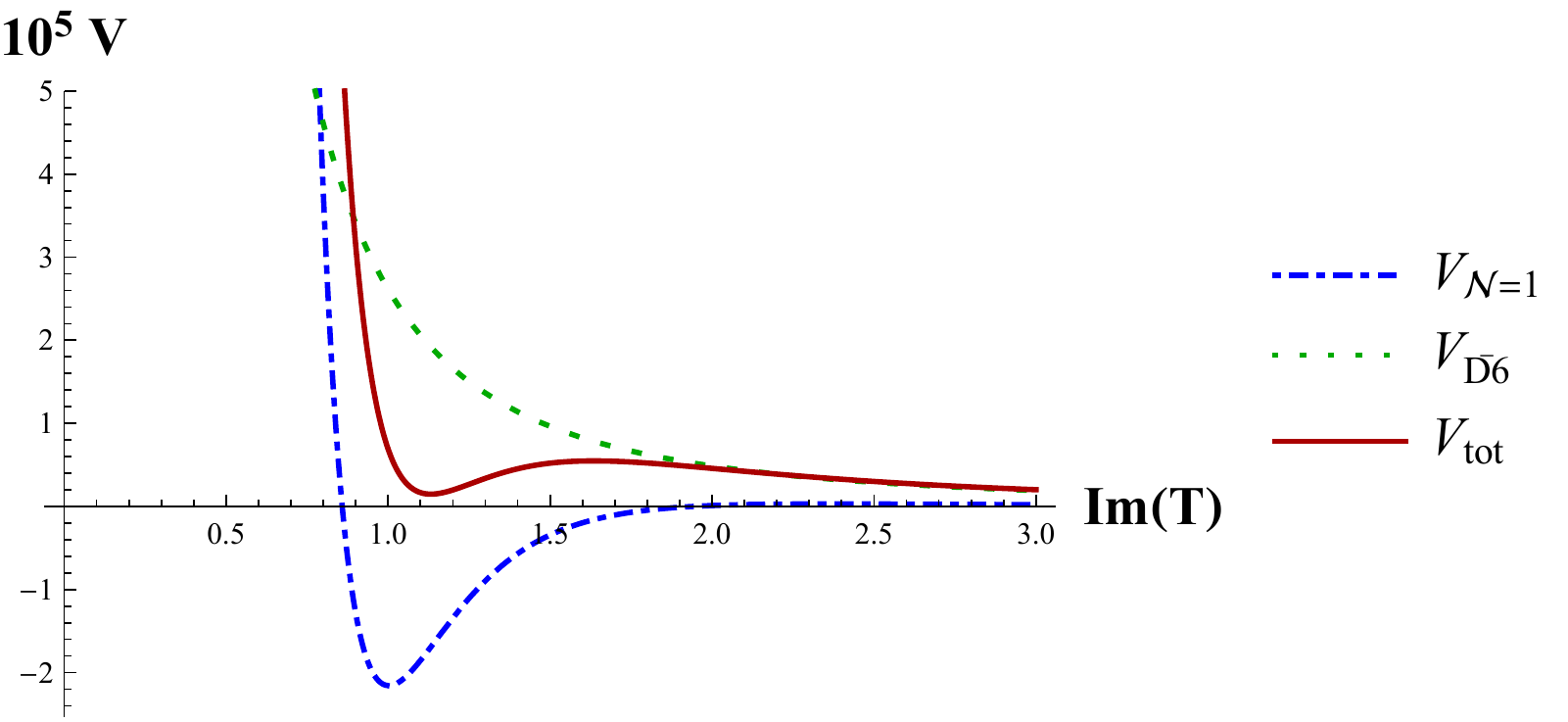}\vspace{15pt}\\
\includegraphics[scale=0.5]{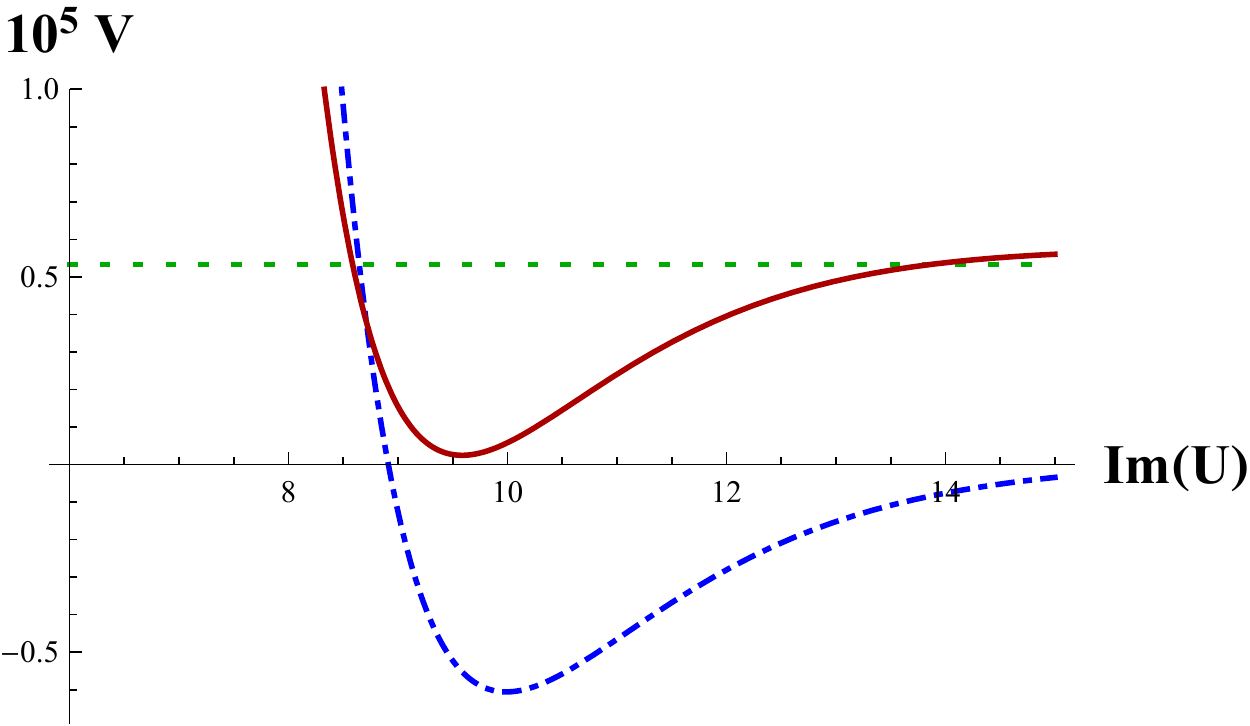}\hspace{10pt} \includegraphics[scale=0.5]{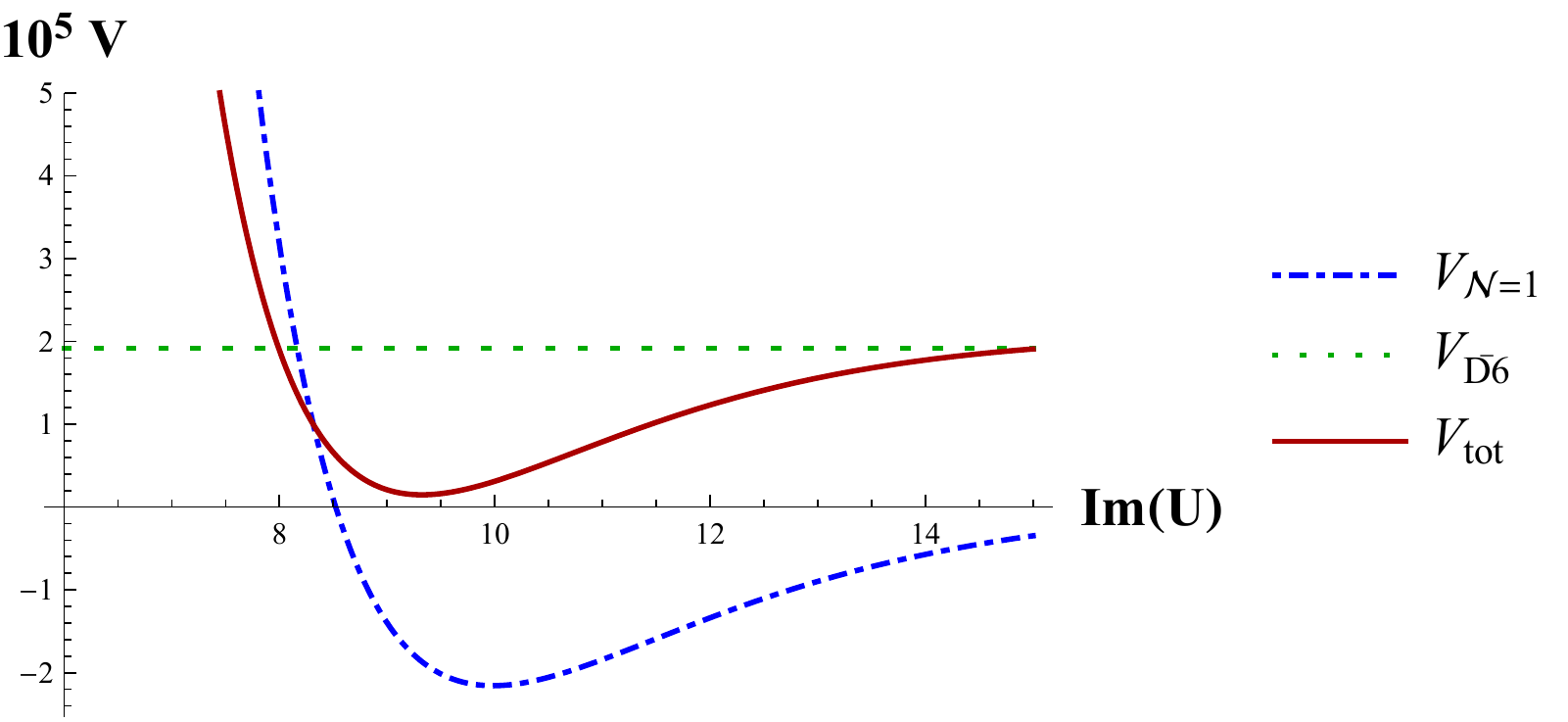}
\caption{2D plots of the total scalar potential $V_{tot}$, the anti-de Sitter potential $V_{\mathcal{N}=1}$ and the $\overline{D6}$ potential $V_{\overline{D6}}$. The left column of plots corresponds to Set 1 while the right is Set 2. Starting from the top we have the Im$(S)$ direction, followed by Im$(T)$ and Im$(U)$. In all plots we see clearly the anti-de Sitter and de Sitter vacua as well as the uplift term.}
\label{fig:3mod2D}
\end{figure}
Therefore, let us add two sets of anti-D6-branes wrapping two 3-cycles in the internal manifold to the setup, as described in section \ref{sec:STUmodel} with the potential presented in equation \rf{eq:vup}. In section \ref{sec:nil} we will explain how such an uplift term can be reproduced from a purely four-dimensional perspective by using a nilpotent chiral goldstino multiplet.

\begin{figure}[H]
\hspace{100pt}
\includegraphics[scale=0.53]{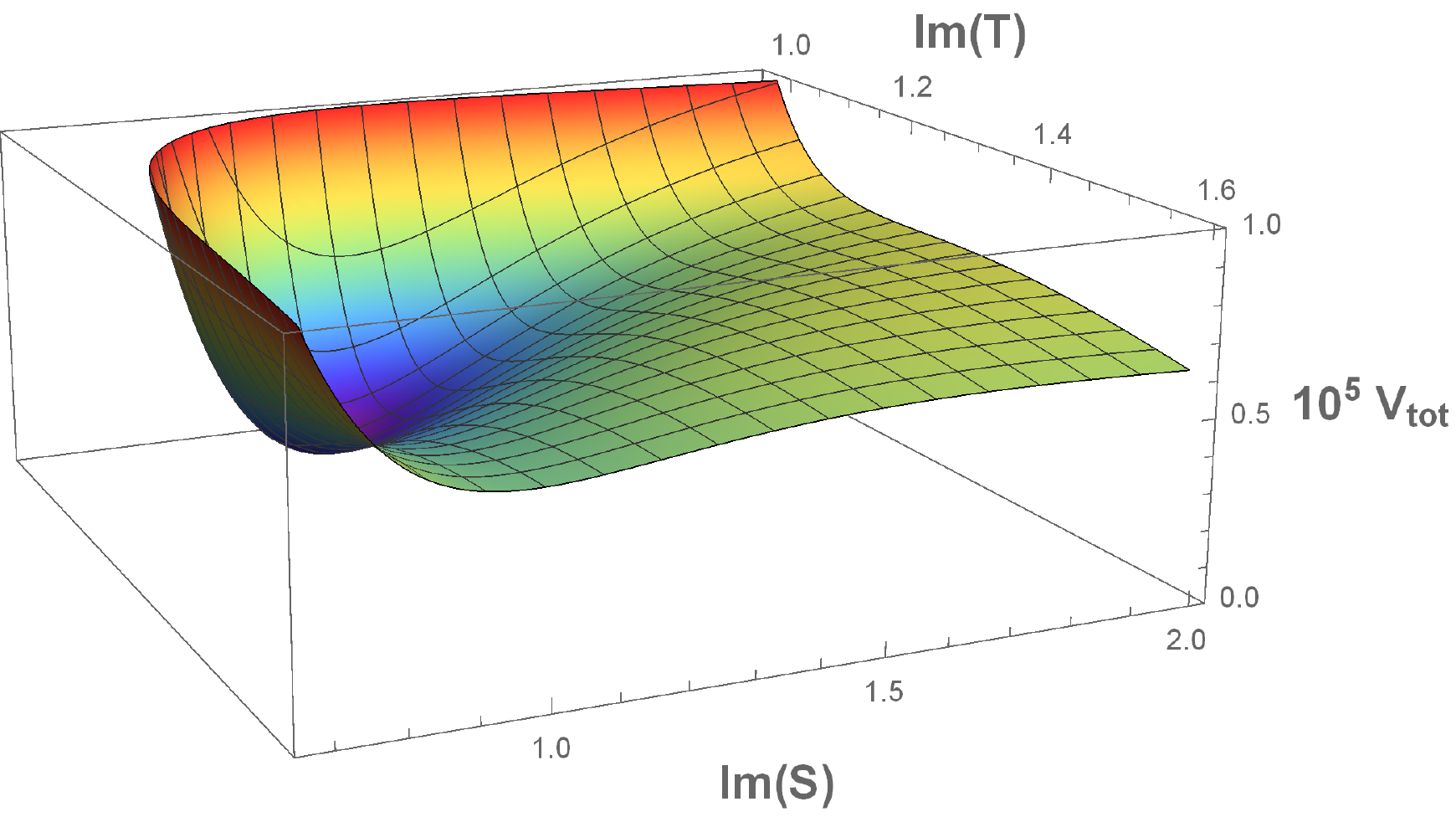}\vspace{15pt}\\
\includegraphics[scale=0.53]{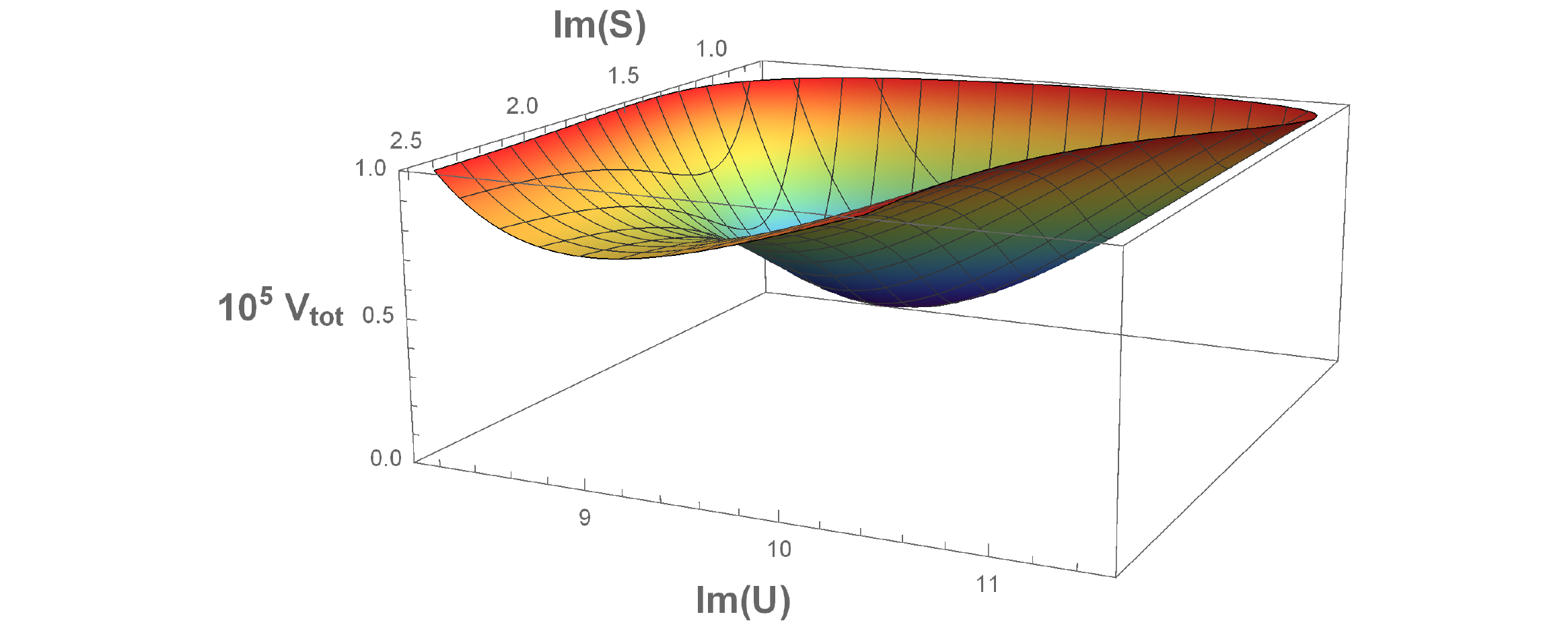}\vspace{15pt}\\
\hbox{\hspace{50pt}\includegraphics[scale=0.53]{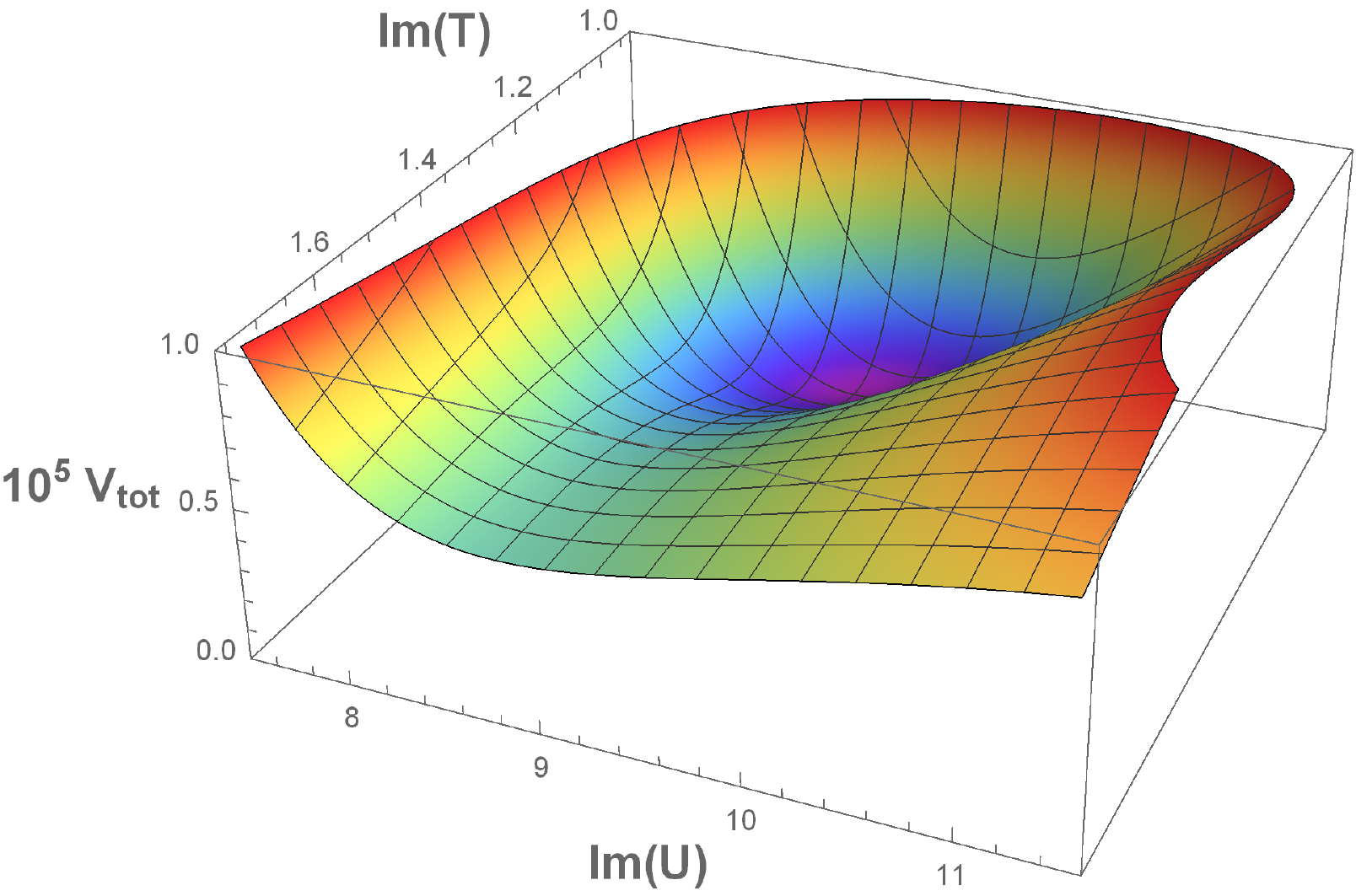}}
\caption{3D plots of the de Sitter potential for the Set 2 of parameters. We have the following slices. Top: Im$(S)$ and Im$(T)$, Middle: Im$(S)$ and Im$(U)$, Bottom: Im$(U)$ and Im$(T)$. In all three different plots the de Sitter minimum is clearly visible and it is metastable.}
\label{fig:3mod3D}
\end{figure}

By appropriately tuning the parameters $\mu_1$ and $\mu_2$, it is possible to obtain de Sitter vacua in which the cosmological constant can be arbitrary small and we could even match it with the measured value at present. However, for convenience of the presentation, we choose the following values:
\be
\begin{aligned}
&\text{Set 1} \qquad \mu_1^4 = 2.01\cdot 10^{-6}, \qquad \mu_2^4=5.21\cdot 10^{-6},\\
&\text{Set 2} \qquad \mu_1^4=\mu_2^4 = 1.34 \cdot 10^{-5}.
\end{aligned}
\ee
After introducing the uplift, we need to check the stability of the resulting de Sitter vacua by calculating the masses of the various fields. The results are reported in table \ref{tab:massesDS}.
\begin{table}[H]
\center
\begin{tabular}{|c|c|c|c|c|c|c|}\hline
& $m_{1}^{\,2}$ & $m_{2}^{\,2}$  & $m_{3}^{\,2}$ & $m_{4}^{\,2}$ & $m_{5}^{\,2}$ & $m_{6}^{\,2}$ \\\hline
Set 1 & $3.43\cdot10^{-4}$& $3.38\cdot 10^{-4}$& $6.46\cdot 10^{-5}$ & $5.40\cdot 10^{-5}$& $4.15 \cdot 10^{-5}$ & $3.47 \cdot 10^{-5}$ \\\hline
Set 2 & $8.00 \cdot 10^{-4}$ & $7.40 \cdot 10^{-4}$ & $1.76\cdot 10^{-4}$ & $1.63 \cdot 10^{-4}$ & $1.61 \cdot 10^{-4}$& $1.50 \cdot 10^{-4}$  \\\hline
\end{tabular}
\caption{The canonically normalized masses squared after the uplift remain positive. This shows that the de Sitter vacua under investigation are metastable.}
\label{tab:massesDS}
\end{table}
In figure \ref{fig:3mod2D}, two-dimensional slices of the scalar potential are shown. The presence of the anti-de Sitter and de Sitter vacua described so far is clearly visible. We notice that the position of the minimum shifts after the uplift. This is expected since the addition of the anti-D6-branes modifies the form of the scalar potential. However, the actual shift is only up to a maximum of about $10\%$, which tells us that the masses are large enough to keep the minimum almost in place. 
In figure \ref{fig:3mod3D}, three-dimensional plots of the scalar potential are shown for our Set 2 of parameters. Again, a metastable de Sitter minimum can clearly be seen.
\label{sec:uplift}
We also checked that the total scalar potential is extremized at the position of the de Sitter vacuum, namely $\partial_i V_{tot}|_\text{min}=0$. We indeed find that this is the case within the limits of our numerical precision.

Finally, we noticed that one can also get metastable de Sitter vacua by including an anti-D6-brane uplift on either one of the two 3-cycles, i.e. by setting $\mu_1=0$ or $\mu_2=0$. It is also possible to use Euclidean D2-instantons instead of gaugino condensation on a stack of D6-branes by setting $a_S=2\pi$ and/or $a_T=2\pi$. Therefore, we see once more that this construction is rather robust and does not depend on any particular choice of parameters.

\subsection{Simplifying or generalizing the setup}

So far, we have described an STU model with three independent moduli. Such a model can be understood as a subcase of a more generic setup, with seven moduli, namely $S$, $T_1$, $T_2$, $T_3$, $U_1$, $U_2$ and $U_3$,  in which we identified two sets of three moduli: $T_1 = T_2 = T_3 \equiv T$ and $U_1 = U_2 = U_3 \equiv U$. This corresponds to, for example, identifying three different tori of the compactification manifold. Along this logic,  we can also set $T = S$ and arrive at the following simplified \K potential and superpotential:
\be
\begin{aligned}
K &= -4\log\left( -\rmi (S - \bar{S})\right) - 3 \log\left( -\rmi(U- \bar{U})\right)\,,\\
W &= f_6 + A_S e^{\rmi a_S S} + A_U e^{\rmi a_U U}\,.
\end{aligned}
\ee
It is now interesting to ask whether or not this also leads to a viable model for a de Sitter uplift with anti-D6-branes. Indeed, it turns out that the construction of the model works out exactly in the same way as described in the previous sections. Even more interestingly, the same statement seems to be true for the generalization to the seven moduli case.

\section{Four-dimensional action with $S$, $T$, $U$ and a nilpotent multiplet}
\label{sec:nil}
As it was first studied in \cite{Kallosh:2018nrk} and as we already mentioned before in the present work, it is possible to include the contribution from the anti-D6-branes to the scalar potential directly in the four-dimensional \K potential and superpotential. This is an example of the general fact that a nilpotent chiral goldstino superfield $X$, that satisfies $X^2 = 0$, can be used to include the contributions from anti-Dp-branes into the potentials.

A chiral multiplet satisfying $X^2=0$ has only one physical degree of freedom: the scalar is in fact given as a fermion bilinear. In superspace notation this means that the chiral superfield $X = \phi + \sqrt{2} \chi \theta + F \theta^2$, with the superspace coordinates $\theta$ and auxiliary field $F$, reduces to $X = \chi^2/(2F) + \sqrt{2} \chi \theta + F \theta^2$ upon enforcing the nilpotent constraint. An important feature of this method is that, after the inclusion of a nilpotent multiplet, supersymmetry will be realized non-linearly. 

The general form of the \K potential and superpotential of four-dimensional $\N=1$ supergravity including a nilpotent chiral multiplet and coming from type IIA string theory is given in equation (35) of \cite{Kallosh:2018nrk} (the nilpotent multiplet is called $S$ there). When specialized to our model, we find:
\be 
\begin{aligned}
\label{EFTX}
K = &- \log\left( -\rmi ( S-\bar{S})\right) - 3 \log\left( - \rmi (T-\bar{T})\right) \\&- \log\left( \left[-\rmi (U - \bar{U})\right]^3 -  \frac{X \bar{X}}{e^{\mathcal{A}_1} N_{\overline{D6}_1} \left(-\rmi (S-\bar{S})\right) +e^{\mathcal{A}_2} N_{\overline{D6}_1}  \left(-\rmi (T-\bar{T})\right)} \right)\,, \\
W = &f_6 + A_S e^{\rmi a_S S} + A_T e^{\rmi a_T T} + A_U e^{\rmi a_U U} + \mu^2 X\,.
\end{aligned}
\ee
Now we can calculate the scalar potential using the same formulas as before and implementing the nilpotency of $X$ at the end. Then, we compare the resulting expression with the one we obtained from the uplift in \eqref{eq:vup}. They turn out to be exactly equal once we identify 
\be
\mu_1^4 = \frac18\mu^4 e^{\mathcal{A}_1} N_{\overline{D6}_1},\qquad\qquad \mu_2^4 = \frac18\mu^4 e^{\mathcal{A}_2} N_{\overline{D6}_2}.
\ee
This procedure has the obvious advantage that everything can be included in the four-dimensional supergravity description and shows once more the usefulness of non-linear supergravity.

\section{ Seven Moduli Model}
\label{sec:7mod}

The STU-model studied so far is actually a simplified version of a more general type IIA model with seven moduli, in which one identifies $T_1 = T_2 = T_3 \equiv T$ and $U_1 = U_2 = U_3 \equiv U$. In this section, we give some details of the analysis we performed on the seven-moduli model. In particular, following the same strategy as in the STU-model and by using anti-D6-branes, we have been able to find again stable de Sitter vacua even in the completely non-isotropy case.

The seven-moduli model, before the uplift, is described by the following  \K potential and superpotential: 
\be
\begin{aligned}
K&=  - \log\left(-\rmi (S-\bar{S})\right) - \sum_{i=1}^3  \log\left(-\rmi (T_i-\bar{T}_i)\right) - \sum_{i=1}^3 \log\left(-\rmi (U_i-\bar{U}_i)\right)\,,\\
W& = f_6 + A_{S}\, e^{\rmi a_S S} +  \sum_{i=1}^3 A_{T,i}\, e^{\rmi a_{T,i} T_i}+ \sum_{i=1}^3 A_{U,i}\, e^{\rmi a_{U,i} U_i}\,.
\end{aligned}
\label{eq:7modpot}
\ee
Once uplifting anti-D6-branes are introduced, they contribute to the scalar potential with the additional term
\be 
\begin{aligned}
V_{\overline{D6}} =\; &\frac{\mu_1^{\,4}}{Im(T_1)  Im(T_2)  Im(T_3)} +\frac{\mu_2^{\,4}}{Im(S)  Im(T_2)  Im(T_3)} \\+&\frac{\mu_3^{\,4}}{Im(S)  Im(T_1)  Im(T_3)} +\frac{\mu_4^{\,4}}{Im(S)  Im(T_1)  Im(T_2)}\,,
\label{eq:7moduplift}
\end{aligned}
\ee 
where $\mu_1$, $\mu_2$, $\mu_3$ and $\mu_4$ are parameters. Even in this case, the anti-D6-brane can be implemented into supergravity by means of a nilpotent chiral multiplet $X$. The resulting effective theory is a straightforward generalization of \eqref{EFTX}.

The analysis of the models \eqref{eq:7modpot} proceeds as usual. First, we find a stable and supersymmetric AdS vacuum, by solving the F-term equations $D_i W=0$. Again, we will give here only two different sets of parameters corresponding to two distinct solutions, but the choice we made is by no means exceptional. The parameters we used are reported in table \ref{tab:7modpara}.

\begin{table}[H]
\center
\begin{tabular}{|c|c|c|c|c|c|c|c|}\hline
      & $S_0$ & $T_{1,0}$ & $T_{2,0}$ & $T_{3,0}$ & $U_{1,0}$ & $U_{2,0}$ & $U_{3,0}$  \\\hline
Set 1 & $1$ & $1$ & $1$ & $1$ & $10$ & $10$ & $10$ \\\hline
Set 2 & $1$ & $1.4$ & $1.5$ & $1.6$ & $11$ & $12$ & $13$ \\\hline\hline
 & $a_S$ & $a_{T_1}$ & $a_{T_2}$ & $a_{T_3}$ & $a_{U_1}$ & $a_{U_2}$ & $a_{U_3}$ \\\hline
Set 1& $3$ & $3$ & $3$ & $3$ & $0.5$ & $0.5$ & $0.5$  \\\hline
Set 2& $3.6$ & $3.7$ & $3.8$ & $3.9$ & $0.33$ & $0.34$ & $0.36$ \\\hline
\end{tabular}
\caption{Two choices of the parameters for the seven-moduli model. In addition, we take $f_6=1$ for Set 1, while $f_6=3/\sqrt 2$ for Set 2.}
\label{tab:7modpara}
\end{table}

The values of the parameters $A_i$ and of the exponentials $e^{-a_i {\rm Im}\Phi_i}$ for these two solutions are reported in the following table.

\begin{table}[H]
\center
\begin{tabular}{|c|c|c|c|c|c|c|c|}\hline& $A_S$ & $A_{T_1}$ & $A_{T_2}$ & $A_{T_3}$ & $A_{U_1}$ & $A_{U_2}$ & $A_{U_3}$ \\\hline
Set 1& $-1.70$ & $-1.70$ & $-1.70$ & $-1.70$ & $-7.55$ & $-7.55$ &$-7.55$   \\\hline
Set 2& $-6.09$ & $-20.6$ & $-31.4$ & $-49.2$ & $-6.22$ & $-8.69$ & $-13.8$\\\hline\hline
& $e^{-a_S {\rm Im}S}$ & $e^{-a_{T_1} {\rm Im}T_1}$ & $e^{-a_{T_2}{\rm Im}T_2}$ & $e^{-a_{T_3} {\rm Im}T_3}$ & $e^{-a_{U_1} {\rm Im}U_1}$ & $e^{-a_{U_2} {\rm Im}U_2}$ & $e^{-a_{U_3} {\rm Im}U_3}$ \\\hline
Set 1& $0.0499$ & $0.0499$ & $0.0499$ & $0.0499$ & $0.00674$ & $0.00674$ &$0.00674$   \\\hline
Set 2& $0.0273$ & $0.00563$ & $0.00335$ & $0.00195$ & $0.0265$ & $0.0169$ & $0.00928$\\\hline
\end{tabular}
\caption{The values of $A_i$ and of $e^{-a_i {\rm Im}\Phi_i}$ for the two anti-de Sitter solutions.}
\label{tab:7modpara}
\end{table}

Then, we can uplift these vacua by introducing anti-D6-branes. We chose the following values for the uplifting parameters.
\be
\begin{aligned}
\text{Set 1: }& \quad \mu_1^{\,4} = 1.23 \cdot 10^{-6}, \quad \mu_2^{\,4} = 1.23 \cdot 10^{-6} ,\quad \mu_3^{\,4} = 3.11 \cdot 10^{-6}, \quad \mu_4^{\,4} = 2.14 \cdot 10^{-6}.\\
\text{Set 2: }& \quad \mu_1^{\,4} = 5.52 \cdot 10^{-6}, \quad \mu_2^{\,4} = 3.45 \cdot 10^{-6}, \quad \mu_3^{\,4} = 3.68 \cdot 10^{-6}, \quad \mu_4^{\,4} = 3.94 \cdot 10^{-6}\,.
\end{aligned}
\end{equation}

After the uplift, we find stable de Sitter vacua. The squared masses for the seven-moduli are reported in table \ref{tab:7modmasses}, for both the de Sitter solutions.

\begin{table}[H]
\center
\begin{tabular}{|c|c|c|}\hline
               & Set 1 & Set 2 \\\hline
$m_1^{\,2}$    &$3.33 \cdot 10^{-4}\,$&$3.95 \cdot 10^{-4}\,$\\\hline
$m_2^{\,2}$    &$3.32 \cdot 10^{-4}$&$3.62 \cdot 10^{-4}$\\\hline
$m_3^{\,2}$    &$1.55 \cdot 10^{-4}$&$1.98 \cdot 10^{-4}$\\\hline
$m_4^{\,2}$    &$1.55 \cdot 10^{-4}$&$1.73 \cdot 10^{-4}$\\\hline
$m_5^{\,2}$    &$1.44 \cdot 10^{-4}$&$1.58 \cdot 10^{-4}$\\\hline
$m_6^{\,2}$    &$1.44 \cdot 10^{-4}$&$1.45 \cdot 10^{-4}$\\\hline
$m_7^{\,2}$    &$6.31 \cdot 10^{-5}$&$1.32 \cdot 10^{-4}$\\\hline
$m_8^{\,2}$    &$4.99 \cdot 10^{-5}$&$1.23 \cdot 10^{-4}$\\\hline
$m_9^{\,2}$    &$4.13 \cdot 10^{-5}$&$1.10 \cdot 10^{-4}$\\\hline
$m_{10}^{\,2}$ &$4.13 \cdot 10^{-5}$&$1.00 \cdot 10^{-4}$\\\hline
$m_{11}^{\,2}$ &$4.11 \cdot 10^{-5}$&$8.56 \cdot 10^{-5}$\\\hline
$m_{12}^{\,2}$ &$3.52 \cdot 10^{-5}$&$7.35 \cdot 10^{-5}$\\\hline
$m_{13}^{\,2}$ &$3.08 \cdot 10^{-5}$&$6.25 \cdot 10^{-5}$\\\hline
$m_{14}^{\,2}$ &$2.85 \cdot 10^{-5}$&$5.50 \cdot 10^{-5}$\\\hline
\end{tabular}
\caption{The canonically normalized masses in the de Sitter vacuum for the  seven-moduli case.}
\label{tab:7modmasses}
\end{table}

Once more, we stress that no particular fine tuning is necessary also for the uplift parameters. To conclude, we have shown that even in this more complicated seven-moduli scenario, the analysis proceeds exactly in the same way as in the STU-model that we discussed in detail in the previous sections.

\section{Discussion}
In the past, type IIA string theory was always viewed as a theory which is difficult to make compatible with cosmology \cite{Kallosh:2006fm, Hertzberg:2007wc, Haque:2008jz, Flauger:2008ad, Danielsson:2009ff, Wrase:2010ew, Shiu:2011zt, Junghans:2016uvg, Andriot:2016xvq, Junghans:2016abx, Andriot:2017jhf}. A great effort in this direction was based on a complicated polynomial in $S$, $T$ and $U$ in the superpotential, as shown in equation \eqref{eq:Wpert}, with four types of fluxes and also terms associated with the curvature of the compact manifold.  Even with all these different contributions, it was not possible to easily produce de Sitter minima in type IIA string theory compactified to four dimensions.

The  first, unexpected, result of this paper is that instead of a complicated polynomial in $S$, $T$ and $U$ in the superpotential, as shown in equation \eqref{eq:Wpert}, one can use just a six-flux and non-perturbative exponential terms, given in equations \eqref{our} and \eqref{np} to stabilize all moduli in a supersymmetric anti-de Sitter minimum with all six mass eigenvalues positive. Thus, with the standard \K potential for the STU model in equation \eqref{our} and with the simple superpotential 
\be
W= f_6+ A_S e^{\rmi a_S S}+ A_T e^{\rmi a_T T}+A_U e^{\rmi a_U U}\,,
\label{new}\ee
it is easy to find parameters which lead to anti-de Sitter minima. 

The second result about the uplifting role of the anti-D6-brane was predicted in \cite{Kallosh:2018nrk}, but not explicitly realized in the examples there. Here, we have found that the stable anti-de Sitter minima obtained in the STU model in equation \eqref{our} are upliftable to stable de Sitter minima, i.e. again we find all six mass matrix eigenvalues to be positive.

This situation has to be contrasted with our earlier efforts, which are reported in the appendix.  In particular, we studied STU models where in addition to six-flux and non-perturbative corrections we engaged also other fluxes and curvature terms, like the ones shown in equation \eqref{eq:Wpert}.  In these setups, we have typically encountered tachyons in the anti-de Sitter extremum, which made these models unsuitable for uplifting. Another class of models with two moduli is presented in appendix \ref{sec:failedmodels}. It was obtained from the STU model after the identification $S=T$. It had the feature that the U-dependence in $W$ was polynomial, but in the S-direction we engaged two exponents, as in \cite{Kallosh:2004yh}. As a result, we were able to find anti-de Sitter minima in these models, however, the anti-D6-brane uplift failed again: we were not able to find a stable de Sitter minimum for the complex $U$ and the complex $S$ directions.
 
In view of these failed examples it might sound puzzling as to why the simple model in \eqref{new} works well, whereas other models with a significant polynomial dependence on the moduli do not work. One explanation is that our new STU model can be qualified as (KKLT)$^3$: we took a KKLT model, which is known to work in the case of one complex modulus, and we did the same with the other directions. Indeed, we just have a constant term and exponents in each direction in the superpotential. We have also learned that the no-scale structure of the \K potential is not really important here. In fact, our STU model leads to de Sitter minima for cases with different contribution to $K$: $- \log (-\rmi (S-\bar S))$,  $- 3 \log (-\rmi (T-\bar T))$ and $  - 3 \log (-\rmi (U-\bar U))$.  

In section \ref{sec:7mod}, we have also checked that the same principle works for the seven-moduli case, namely (KKLT)$^7$, with
\be
K=  - \sum_{i=1} ^{7}\log (-\rmi (\Phi_i-\bar \Phi_i))\,, \qquad W= f_6 +\sum_{i=1} ^{7}  A_i e^{\rmi a_i \Phi_i}\,.
\ee
As we explained before, our STU model in equation \rf{our} corresponds to the seven-moduli case  where $T_1=T_2= T_3$ and $U_1=U_2= U_3$ are identified. Such a seven-moduli setup is particularly interesting with regard to the CMB B-mode targets \cite{Ferrara:2016fwe,Kallosh:2017ced}. 

It would also be interesting to go beyond the simple STU model and study some other compactifications. In particular, the original GKP solution in type IIB \cite{Giddings:2001yu} has been T-dualized to find type IIA no-scale Minkowski vacua, in which some of the moduli are stabilized by fluxes \cite{Kachru:2002sk, Grana:2006kf, Andriolo:2018yrz}. These solutions have a build in tadpole cancellation condition and provide a natural starting point for the construction of dS vacua following the original KKLT approach, that we adapted and generalized here in the type IIA setting.

\section*{Acknowledgement}
We are grateful to R. Blumenhagen, S. Kachru, A. Linde and T. Van Riet for important discussions. RK is supported by SITP and by the US National Science Foundation Grant  PHY-1720397, and by the  Simons Foundation Origins of the Universe program (Modern Inflationary Cosmology collaboration),  and by the Simons Fellowship in Theoretical Physics.
The work of NC, CR and TW is supported by an FWF grant with the number P 30265.
NC and CR are grateful to  SITP for the hospitality while this work was performed. CR is furthermore grateful to the Austrian Marshall Plan Foundation for making his stay at SITP possible.

\appendix

\section{How fluxes prohibit stable solutions}
\label{appA}
So far, in the literature (see for example \cite{Marchesano:2019hfb} and references therein), tree-level contributions were usually employed in order to find de Sitter vacua, while in the present work we studied the opposite situation, which made the task much easier. For completeness, in this appendix we want to describe some models we investigated but that ultimately did not allow for a consistent uplift to de Sitter. 

Our starting point was the same \K  potential we used in the main part of this work, namely
\be
K = - \log \left(-\rmi (S-\bar S)\right) - 3 \log \left(-\rmi (T-\bar T)\right)-3 \log \left(-\rmi (U-\bar U)\right)\,,
\ee
and the more general superpotential $W=W_{pert}+W_{np}$ including flux parameters not only for 6-form flux, but also 4-, 2- and 0-form flux, as well as non-perturbative corrections:
\be
\begin{aligned}
W_{pert} &= f_6 + f_4 U + f_2 U^2 + f_0 U^3 + \left( h_T + r_T U\right) T + \left( h_S + r_S U\right) S \qquad\text{and}\\
W_{np}  &=  A_S e^{\rmi a_S S} + A_T e^{\rmi a_T T} + A_U e^{\rmi a_U U}\, .
\end{aligned}
\ee
Again, we assume that all parameters, $f_i$, $i=0, 2, 4, 6$, $A_S, A_T, A_U$ and $a_S, a_T, a_U$, are real.

Starting from these $K$ and $W$ we have a large amount of potential models. Indeed, our choices include identifying moduli, setting flux parameters and/or non-perturbative corrections to zero. It is important to note that one would expect that the non-perturbative contributions should not be significant when compared to the tree level in $W_{pert}$. This means we should set $A_i = 0$, $i = \{S, T, U\}$, if the corresponding modulus appears at tree level.

\subsection{STU models with fluxes}
If one follows the logic concerning the coexistence of tree level and non-perturbative contributions lined out above, it seems that one generally arrives at an anti-de Sitter critical point that includes at least one tachyon. Indeed, while tuning the parameters does allow to modify the masses, we were not able to get rid of all tachyons: at least one mass remained negative. 
On one hand we were aware of the well-known no-go theorems in \cite{Hertzberg:2007wc, Haque:2008jz, Flauger:2008ad}, that tell us that at least $f_0$ and $r_S$ or $r_T$ should be non-vanishing in order to get a de Sitter vacuum. On the other hand, since we evaded the no-go's by including non-perturbative corrections, it is reasonable to consider models that do not obey these conditions.

Since we were not able to arrive at a stable anti-de Sitter vacuum with all masses positive, there is little hope that the uplift will lead to a stable de Sitter vacuum and thus we consider this class of models to be not viable.

\subsection{A KL-type of model}

In \cite{Kallosh:2004yh} a modification of the non-perturbative terms was proposed where one includes two different exponentials for the same modulus. For a general modulus $Y$ this looks like
\be 
A_{Y} e^{\rmi a_Y Y} - B_{Y} e^{\rmi b_Y Y}\,
\ee
and it often improves the situation when the requirement that $ e^{- a_Y \text{Im}(Y)}$ needs to be small is in the way of, for example, positive masses.

In fact in a model where we identified $T = S$ and introduced two exponentials in the $S$-direction, meaning we had
\be
\begin{aligned}
K &= -4 \log\left( -\rmi (S - \bar{S})\right) - 3 \log\left( -\rmi(U- \bar{U})\right)\,,\\
W &= f_6 + f_4 U + f_2 U^2 + f_0 U^3 + A_S e^{\rmi a_S S} - B_S e^{\rmi a_S S}\,,
\end{aligned}
\ee
we were able to find a supersymmetric, stable anti-de Sitter minimum with all masses positive. Interestingly, solving for $D_iW = 0$ actually sets $f_0 = 0$ and no solution exists where this is not satisfied.

\label{sec:failedmodels}
\begin{figure}[H]
\begin{center}
\includegraphics[scale=0.37]{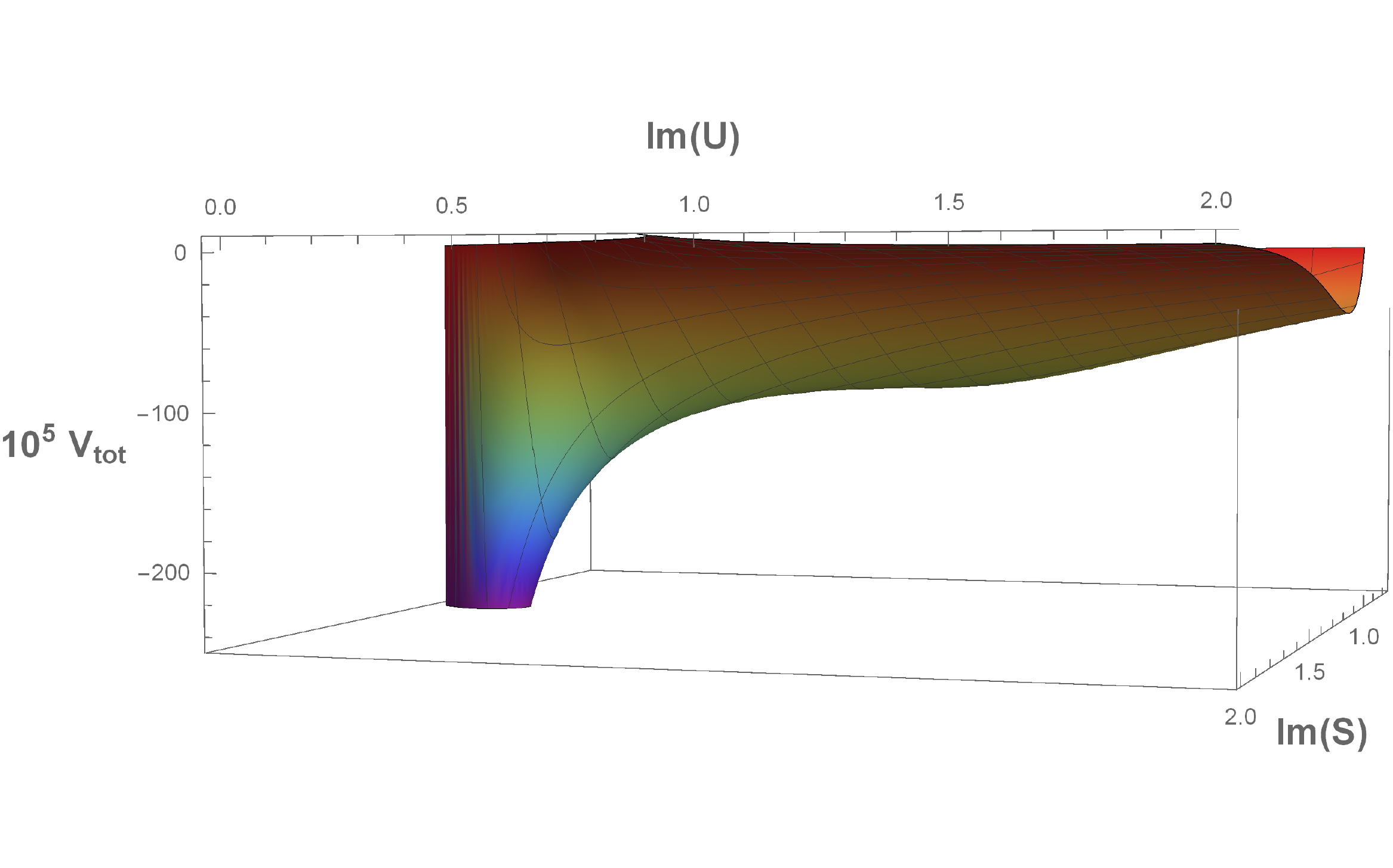}\vspace{15pt}\\
\includegraphics[scale=0.37 ]{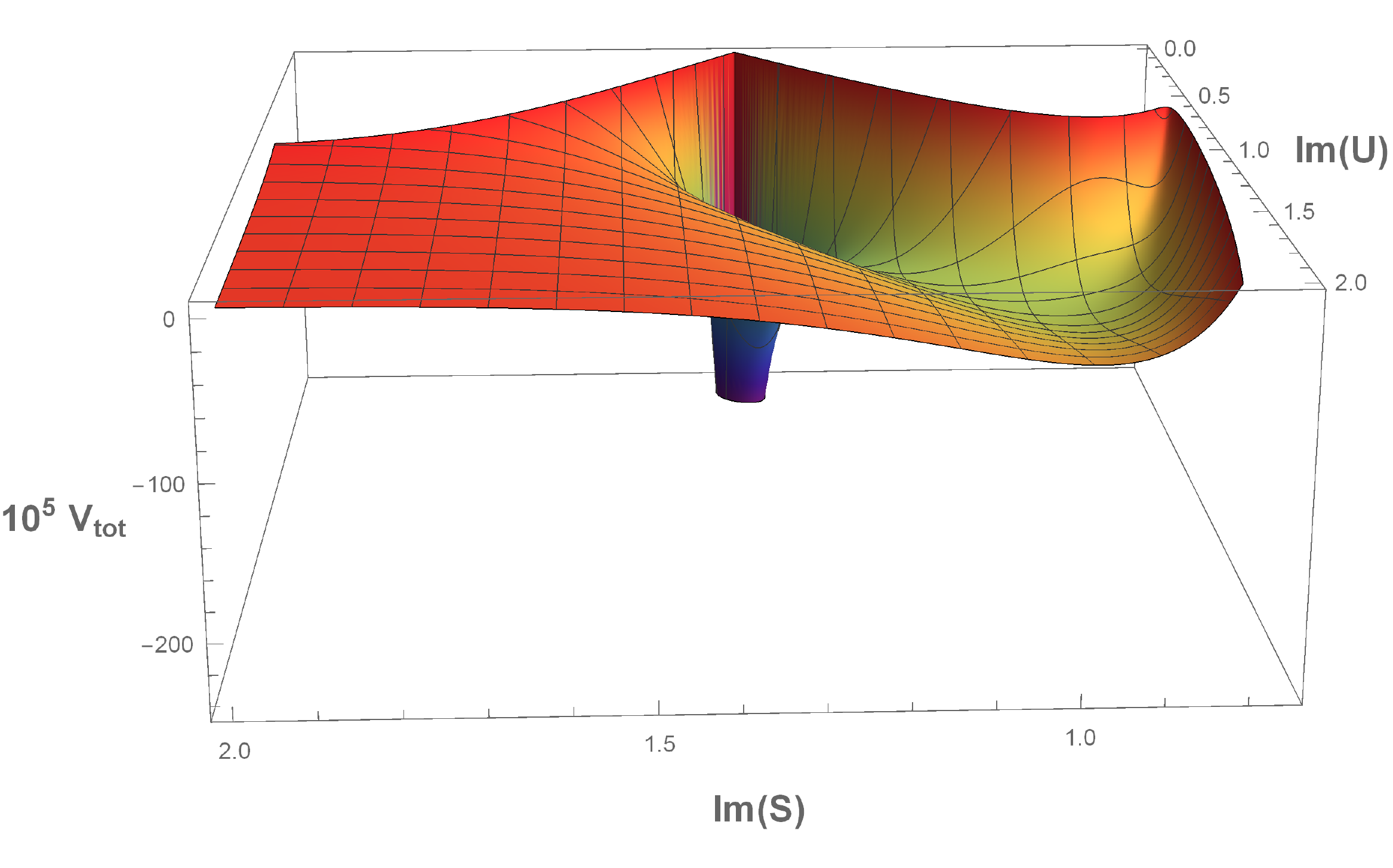}
\end{center}
\caption{The 3D plot for the anti-de Sitter scalar potential does exhibit a stable minimum. However, after the uplift using a anti-D6-brane this will lead to a runaway.}
\label{fig:2modKL}
\end{figure}

Unlike for what happens in our working models presented in the main text, finding all masses positive is not an easy task in this setup. Instead, we needed to tune the parameters $a_S$ and $b_S$ quite a bit. Indeed, this was our reason to include two exponentials: only then were large values for $a_S$Im$(S)$ and $b_S$Im$(S)$ possible, which is a necessary requirement in order such that higher order non-perturbative corrections can be neglected.

After introducing the uplift, as outlined in section \ref{sec:uplift}, one is unable to find a stable de Sitter vacuum. In figure \ref{fig:2modKL} we show the anti-de Sitter scalar potential before the uplift. After the uplift, in all examples which we studied, this leads to a runaway and no stable de Sitter solution.

\bibliographystyle{JHEP}
\bibliography{lindekalloshrefs}
\end{document}